\documentclass{article}

\usepackage{arxiv}

\usepackage{graphicx}% Include figure files
\usepackage{dcolumn}% Align table columns on decimal point
\usepackage{bm}% bold math
\usepackage{here}
\usepackage{amsmath,amssymb}

\title{Spatial and Temporal Taylor's Law in 1-Dim Chaotic Maps}

\author{
 Hiroki Kojima \\
  University of Tokyo\\
  \texttt{kojima@sacral.c.u-tokyo.ac.jp} \\
   \And
 Yuzuru Mitsui \\
  University of Tokyo\\
  \texttt{mitsui@sacral.c.u-tokyo.ac.jp} \\
  \And
 Takashi Ikegami \\
  University of Tokyo\\
  \texttt{ikeg@sacral.c.u-tokyo.ac.jp} \\
}

\begin{document}
\maketitle
\begin{abstract}
% We investigate the relationship between the sample mean and variance, known as Taylor's law (TL), from trajectories generated from simple low dimensional chaotic maps. The purpose here is to reveal the relationship between TL of spatial ensembles (STL) and temporal ensembles (TTL). We confirm that the spatial ensemble is equivalent to the iid sampling from the stationary distribution, and STL is explained by the skewness of the distribution. We extend this to the ensembles of dependent variables and show that the difference between the TTL and STL is formally related to the temporal correlation structure. We further investigate the TTL of each map and analytically show the quadratic relationship of variance with the mean, i.e., Bartlett's law, in the case of logistic and tent maps. The TTL of the Hassell model is well-explained by the chunk structure of the trajectories, but, unexpectedly, the TTL of the Ricker model, which has a similar structure, cannot be explained by the chunk structure, and we found that the specific mathematical form of the map decorrelates the mean and variance. These results indicate that temporal correlations is another possible origin of TL.
By using low-dimensional chaos maps, the power law relationship established between the sample mean and variance called Taylor's Law (TL) is studied. In particular, we aim to clarify the relationship between TL from the spatial ensemble (STL) and the temporal ensemble (TTL). Since the spatial ensemble corresponds to independent sampling from a stationary distribution, we confirm that STL is explained by the skewness of the distribution. The difference between TTL and STL is shown to be originated in the temporal correlation of a dynamics. In case of logistic and tent maps, the quadratic relationship in the mean and variance, called Bartlett's law, is found analytically. On the other hand, TTL in the Hassell model can be well explained by the chunk structure of the trajectory, whereas the TTL of the Ricker model have a different mechanism originated from the specific form of the map.

\end{abstract}

% keywords can be removed

\section{Introduction}
We challenged the unexplained widespread power-law behavior between the mean and variance, called Taylor's law (TL) \cite{Taylor1961}, with one-dimensional (1-D) chaotic maps. 
TL was originally found in the field of population ecology and has since been reported more widely, ranging from demographic ecology to prime number distribution \cite{Cohen2016}, complex networks \cite{DeMenezes2004}, and so on \cite{Eisler2008, Taylor2019}. TL expresses that a power-law scaling relationship holds between the mean ($M$) and variance ($V$) as $V = \alpha M^{\beta}$. When the mean and variance are calculated by randomly sampling from a stationary state, it is called spatial TL (STL), and, when it is sampled from a time series, it is called temporal TL (TTL). 

STL and TTL have been analyzed in several dynamical systems of ecological models \cite{Kilpatrick2003,Ballantyne2005,Perry1994,Cohen2013TPB}. For example, Ballantyne \cite{Ballantyne2005} showed that the exponent of TTL becomes 2 when the solution of the map is linearly scaled by changes in the parameters, and Kilpatrick and Ives \cite{Kilpatrick2003} used the noisy Ricker model of multiple species to show TTL between the mean and variance. Perry \cite{Perry1994} found STL in the chaotic regions of the Hassell model controlled by noise, but they provided no analytical explanations. Cohen showed that the exponential growth model satisfies STL with the exponent $\beta = 2$ in the limit of large time \cite{Cohen2013TPB}. 

In this letter, we investigate both STL and TTL in the chaotic regime of 1-D maps, discussing a possible mechanism of both STL and TTL. There are only a few studies that have investigated the relationship of STL and TTL  \cite{Woiwod1982,Saitoh2020,Zhao2019}. They use empirical data with a probabilistic model, but no theoretical explanation was provided. 
We discuss that spatial TL is explained by the skewed distribution function, while temporal TL is dependent on the temporal correlations in a time series.
Zhao et al., \cite{Zhao2019} argued that the skewness of the population abundance was an important factor of the exponent $\beta$. Although they recognized that the autocorrelation of the time series was important, their main numerical simulations did not consider it. Here, we will show that autocorrelation greatly influenced TL.

\section{TL in 1-D chaotic maps}

We examined four discrete one-dimensional chaotic map systems that basically adopted ecological modelings—i.e., logistic map, tent map, Hassell model \cite{Hassell1974, Hasselletal1976} and Ricker model \cite{Ricker1954}—which have been widely used as population models \cite{may1976bifurcations, Murray2007}.

% We examined four discrete 1-dimensional chaotic map systems that basically adopted in ecological modelings: logistic map, tent map, Hassell model \cite{Hassell1974, Hasselletal1976} and Ricker model \cite{Ricker1954}, which have been widely used as population models \cite{may1976bifurcations, Murray2007}.
\begin{equation}
    x_{n+1} = r x_{n} (1-x_{n}),
\label{eq:logistic}
\end{equation}
\begin{equation}
x_{n+1} =\dfrac{\mu}{2} - \mu \left| x_n - \dfrac{1}{2} \right|,
\label{eq:tent}
\end{equation}
\begin{equation}
  x_{n+1} = \dfrac{\lambda x_{n}}{(1 + \kappa x_{n})^{\nu}},
    \label{eq:Hassell}
\end{equation}
\begin{equation}
    x_{n+1} = x_{n}\exp \{r(1-x_{n})\}.
    \label{eq:Ricker}
\end{equation}

\begin{figure}
 \begin{center}
  \includegraphics[width=86mm,height=70mm]{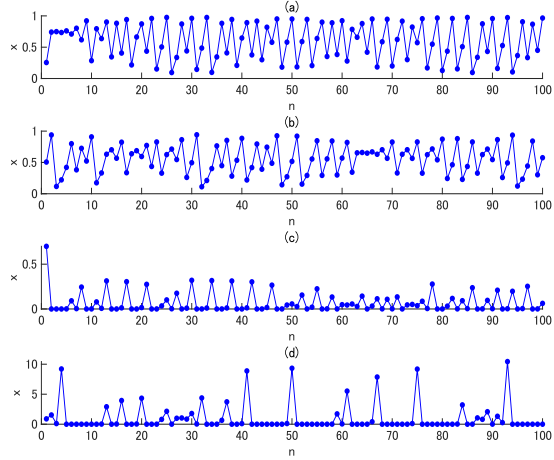}
 \end{center}
 \caption{The chaotic time series of maps used here. (a) Logistic map with $r=3.9$. (b) Tent map with $\mu = 1.9$. (c) Hassell model with $\kappa=10$, $\nu=12$, $\lambda=100$. (d) Ricker model with $r=5$.}
 \label{fig:orbits}
\end{figure}

\noindent The typical time series generated from these maps are shown in FIG. \ref{fig:orbits}.

% We calculated the multiple trajectories by varying the initial states  with fixed parameters. Each trajectory starts from different initial values randomly sampled from a uniform distribution. The length of each trajectory is set at 20,000 time steps. We eliminated the initial transients (discarded the first 10,000 steps) and analyzed the last 10,000 steps ($T=10000$). 
% By denoting the $n$th value of the $i$th trajectory as $x^{i}_{n}$, we analyzed the ensemble of N samples; computing

\begin{figure}
 \begin{center}
  \includegraphics[width=86mm,height=130mm]{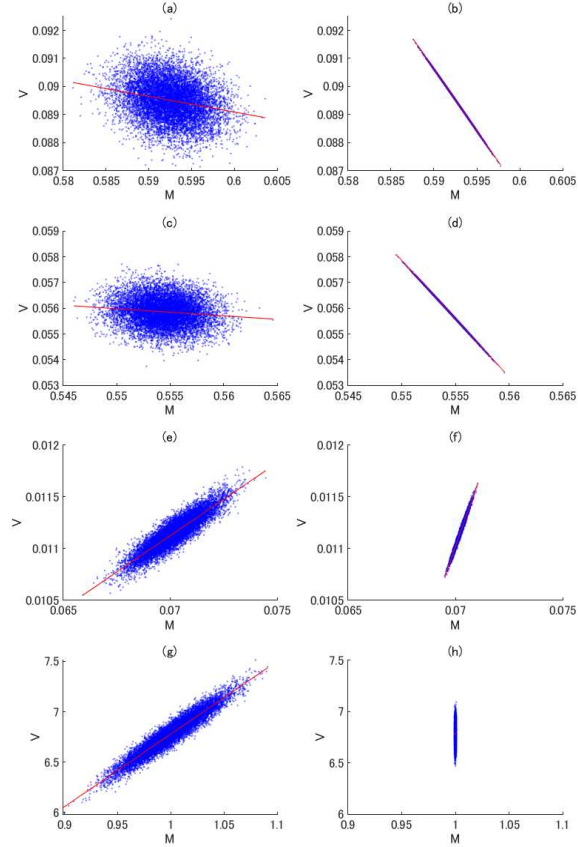}
 \end{center}
 \caption{Results of the numerical simulations of 10,000 sets of $(M, V)$ on the $M-V$ plane. The data points are represented as blue dots, and the linear regression lines are shown as red lines. (a) STL, logistic map with $r = 3.9$. (b) TTL, logistic map with $r = 3.9$. (c) STL, tent map with $\mu = 1.9$. (d) TTL, tent map with $\mu = 1.9$. (e) STL, Hassell model with $\kappa = 10$, $\nu = 12$, $\lambda = 100$. (f) TTL, Hassell model with $\kappa = 10$, $\nu = 12$, $\lambda = 100$. (g) STL, Ricker model with $r = 5$. (h) TTL, Ricker model with $r = 5$.}
 \label{fig:all}
\end{figure}

We calculated the multiple trajectories by varying the initial states with fixed parameters. Each trajectory starts from different initial values randomly sampled from a uniform distribution. The length of each trajectory is set at 20,000 time steps. We eliminated the initial transients (discarded the first 10,000 steps) and analyzed the last 10,000 steps ($T = 10000$). 
By denoting the $n$th value of the $i$th trajectory as $x^{i}_{n}$, we analyzed the ensemble of $N$ samples; computing $M_n = (1/N)\sum_{i=1}^{N} {x^i_n}$ and $V_n = (1/N)\sum_{i=1}^{N} ({x^i_n}-M_n)^2$ for STL. 
As for computing TTL, we compute $M_i = (1/T)\sum_{n=1}^{T} {x^i_n}$ and $V_i = (1/T)\sum_{n=1}^{T} ({x^i_n}-M_i)^2$  (FIG. \ref{fig:all}).

The calculation settings of STL and TTL here can be interpreted as the observation of population density at independent multiple places of the equal environmental conditions (e.g., \cite{Perry1994, Cohen2013PRSB, Cohen2014TE}).

%STL can be interpreted as the observation of population density at independent multiple places of the equal environmental conditions (e.g., \cite{Perry1994, Cohen2013PRSB, Cohen2014TE}).

% STL can be interpreted as the observation of population density at independent multiple places of the equal environmental conditions (see e.g. \cite{Perry1994, Cohen2013PRSB, Cohen2014TE}).

We study the TL relationship in terms of four values as follows: $b = \mbox{cov}(M,V)/\mbox{var}(M)$, $\mbox{var}(M)$, and $\mbox{var}(V)$. The first quantity $b$ corresponds to the slope of the linear regression line on the $M-V$ plane. When the linear regression is valid, this exponent $b$ approximates the exponent $\beta$ of TL as $\beta \simeq \dfrac{\overline{M}}{\overline{V}}b$ ($\overline{M}$ and $\overline{V}$ are the sample mean and variance calculated over all trajectories, respectively) \cite{Cohen2015}.

These results are summarized in the "Numerical" columns in TABLE \ref{tab:results}.

% We study the TL relationship in terms of four quantities as follows: $b = \mbox{cov}(M,V)/\mbox{var}(M)$, $\mbox{var}(M)$ and  $\mbox{var}(V)$. The first quantity $b$ corresponds to the slope of the linear regression line on the M-V plane. When the linear regression is valid, this exponent $b$ approximates the exponent $\beta$ of TL as $\beta \simeq \dfrac{\overline{M}}{\overline{V}}b$ ($\overline{M}$ and $\overline{V}$ are the sample mean and variance calculated over all trajectories) \cite{Cohen2015}.

% Those results are summarized in the "Numerical" columns in TABLE \ref{tab:results}.

\section{The Origins of Spatial TL and Temporal TL}
In the following sections, we will theoretically analyze and reproduce the above arguments.

\subsection{Spatial ensemble}

The spatial ensemble is organized by randomly sampling from a stationary distribution of a given map. The following quantities are given in \cite{cho2005variance,Zhang2007,Cohen2015}:

% The spatial ensemble is organized by randomly sampling from a stationary distribution of a given map. The following quantities %$\mbox{var}(M)$, $\mbox{var}(V)$ and $\mbox{cov}(M,V)$ 
% are given in \cite{cho2005variance,Zhang2007,Cohen2015};

\begin{equation}
  \mbox{var}(M) = \dfrac{\mu_2}{N},
    \label{eq:sample mean}
\end{equation}

\begin{equation}
  \mbox{var}(V) = \dfrac{1}{N}\left(\mu_4 - \dfrac{N-3}{N-1}\mu_2^2 \right),
    \label{eq:sample variance}
\end{equation}

\begin{equation}
  \mbox{cov}(M,V) = \dfrac{\mu_3}{N},
    \label{eq:skewness}
\end{equation}

 \noindent where $N$ is the sample size. $\mu_2$ is equal to the variance (or the second central moment) of the stationary distribution. $\mu_3$ denotes the skewness of the stationary distribution (or the third central moment), and $\mu_4$ denotes fourth moments of the stationary distribution.

%  Here, $\mu_3$ provides the skewness of the stationary distribution, which is measured by $\mu_3= (x-M)^3$ and $\mu_4$ (they are the third and fourth moments of the stationary distribution. ) 
The covariance of the sample mean and variance are proportional to the skewness of the distribution. Cohen et al. \cite{Cohen2015} argued that it is one of the possible origins of TL.

In order to confirm that STL is explained from the skewness of the stationary distribution, we computed a stationary distribution from the 10,000 trajectories to calculate $\mu_2,\mu_3$ and $\mu_4$. The predicted value matches the value of the numerical calculation, as shown in TABLE \ref{tab:results}.

%  where $N$ is the sample size. $\mu_2$ is equal to the variance (or second central moment) of the stationary distribution. Here $\mu_3$ provides the skewness of the stationary distribution, which is measured by  $\mu_3= (x-M)^3$ and $\mu_4$ (they are the third and fourth moment of the stationary distribution. ) 
% The covariance of the sample mean and variance are proportional to the skewness of the distribution. Cohen et al. \cite{Cohen2015} argued that it is one of the possible origins of TL.

% In order to confirm that STL is explained from the skewness, we computed a stationary distribution from the 10,000 trajectories to calculate $\mu_2,\mu_3$ and $\mu_4$. The predicted value matches the value of the numerical calculation as we see in TABLE I.

\subsection{Temporal ensemble}

TL relationships have different appearances in spatial and temporal ensembles. If each trajectory is independent, spatial TL only depends on the stationary distribution, as stated above, but temporal TL also depends on the temporal structure of the trajectories. We first show how temporal correlation in trajectories relates to temporal TL; next, we estimate the temporal TL of each map based on the characteristics of each of the temporal structures.

% TL relationships have different outlooks in spatial and temporal ensembles. If each trajectory is independent, spatial TL only depends on the stationary distribution as we stated above, but temporal TL also depends on the temporal structure of the trajectories. We first show how temporal correlation in trajectories relate to temporal TL, and next, we estimate the temporal TL of each map based on the characteristics of the each temporal structures.

\subsubsection{A Relationship between temporal TL and spatial TL}

Eq.\ref{eq:sample mean}-\ref{eq:skewness} only hold when the variables are sampled independently.
This holds true for the spatial ensembles but not for the temporal ensembles, as each system has its own characteristic memory length.  

The following generalized equations are applicable when the sample size $N$ is large enough (see Appendix):

% Eq.\ref{eq:sample mean}-\ref{eq:skewness} only hold when variables are sampled independently.
% % randomly and uniformly from the ensemble. 
% This holds true for the spatial ensembles, but not for the temporal ensembles, as each system has the characteristic memory length.  

% The following generalized equations are applicable when the sample size $N$ is large enough (\ref:supp);

\begin{eqnarray*}
  \mbox{var}(M) &=& \dfrac{\mu_2}{N} + \dfrac{1}{N^2}\sum_{\tau = 1}^{N-1}2(N-\tau) R_1(\tau) \\
  \mbox{var}(V) &\simeq&  \dfrac{1}{N}\left(\mu_4 - \mu_2^2 \right) + \dfrac{1}{N^2}\sum_{\tau = 1}^{N-1}2(N-\tau) R_2(\tau)\\
  \mbox{cov}(M,V) &\simeq&  \dfrac{\mu_3}{N} + \dfrac{1}{N^2}\sum_{\tau = 1}^{N-1}(N-\tau)(R_{12}(\tau) + R_{12}(-\tau)),
\end{eqnarray*}
 where the first term is equal to Eq.\ref{eq:sample mean}-\ref{eq:skewness} when the system size is large.  The second order terms are given by the following terms. 
\begin{eqnarray*}
    y_i &=& x_i - E\left[x \right] \\
    % \mu_1 &=& \mbox{mean of the stationary distribution} \\ 
    R_1(\tau) &=&  E\left[ y_i y_{i+\tau} \right], \\
    R_2(\tau) &=&  E\left[(y_i^2 - \mu_2)(y_{i+\tau}^2 - \mu_2)\right], \\
    R_{12}(\tau) &=&  E\left[y_i y_{i+\tau}^2 \right],
\end{eqnarray*}

The second order terms will vanish when the variables are uncorrelated. Specifically, the difference between the \mbox{var}($M$) values calculated from the spatial ensembles and temporal ensembles is proportional to the sum of the auto-correlation function.

We can calculate the TTL if we know $R_1(\tau)$, $R_2(\tau)$ and $R_{12}(\tau)$, but, typically, we need to actually sample the trajectories and directly calculate from them to acquire the functions. Below, we estimate the TTL of each map only using map and conspicuous temporal structure information.

% The second order terms will vanish When the variables are uncorrelated. Especially, the difference between \mbox{var}($M$) calculated from spatial ensembles and from temporal ensembles is proportional to the sum of auto-correlation function.

% We can calculate the TTL if we know $R_1(\tau)$, $R_2(\tau)$, and $R_{12}(\tau)$, but usually we need to actually sample the trajectories and directly calculate from the trajectories to acquire the functions. Below, we estimate the TTL of each map only using the information of the map and conspicuous temporal structures.
\subsubsection{Logistic map}

The mean and variance of the temporal ensembles from the logistic map are strongly correlated (FIG. \ref{fig:all} (b)). We found that this is a direct consequence of the quadratic form of the logistic map (Eq.\ref{eq:logistic}). By taking the temporal summation of Eq.\ref{eq:logistic}, we analytically derived the following equation (see Appendix):

% The mean and variance of the temporal ensembles from the logistic map are strongly correlated  (FIG. \ref{fig:all} (b)). We found that this is a direct consequence from the quadratic form of the logistic map (Eq.\ref{eq:logistic}). By taking the temporal summation of Eq.\ref{eq:logistic}, we analytically derived the following equation:[\ref:supp]

\begin{equation}
 V = (1-\dfrac1r) M - M^2.
     \label{eq:logisticVM}
\end{equation}

In general, a dynamic system containing a quadratic function ($ x_ {n + 1} = cx_ {n} ^2 + dx_ {n} $) shows this relationship in the temporal ensemble as follows (see Appendix):

% In general, a dynamic system containing a quadratic function ($ x_ {n + 1} = cx_ {n} ^ 2 + dx_ {n} $) shows this relationship in the temporal ensemble as follows: [\ref:supp]

\begin{equation}
 V = \dfrac{1-d}{c} M - M^2.
     \label{eq:generalQuad}
\end{equation}

% Therefore, when we plot the temporal mean and variance in the logistic map, we can see the curve structure as shown in FIG. \ref{fig:logistic_curve}.

\begin{figure}[H]
 \begin{center}
  \includegraphics[width=86mm]{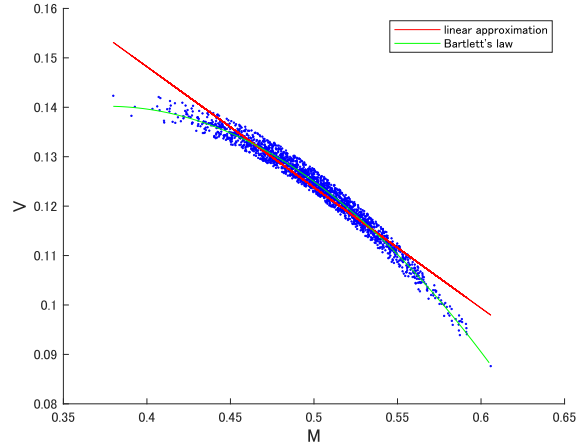}
 \end{center}
 \caption{The typical temporal $M-V$ relationship in the logistic map. Here, $r = 4$, the number of time steps to calculate the temporal mean and variance is 100, and the number of plots is 2,000. The red and green lines represent the linear and Bartlett's law approximations, respectively.}
 \label{fig:logistic_curve}
\end{figure}

This quadratic component is evident when we plot the temporal mean and variance calculated from short trajectories as shown in FIG. \ref{fig:logistic_curve}.

In order to check whether it reproduces the results in TABLE \ref{tab:results}, we calculated the exponent $b$ using $\left. \dfrac{dV}{dM}\right|_{M=\overline{M}}$ ($\overline{M}$:sample mean of the all values in the 10,000 $\times$ 10,000 data sets). The results are shown in the column "Prediction" in TABLE \ref{tab:results}, and they are well fitted with those in the "Numerical" column.

% In order to check whether it reproduces the results in TABLE \ref{tab:results}, we calculated the exponent $b$ by $\left. \dfrac{dV}{dM}\right|_{M=\overline{M}}$ ($\overline{M}$:sample mean of the all values in the 10,000 $\times$ 10,000 data sets). The results are shown in the column of "Prediction" in the TABLE \ref{tab:results} which are well fitted with those in "Numerical" column.

\subsubsection{Tent map}
The TTL of the tent map can be also analytically derived from the equation of the map.
\begin{equation}
 V = \dfrac{\mu}{\mu + 1}M - M^2.
         \label{eq:tentVM}
\end{equation}

This quadratic relation is clearly seen when we plot the temporal mean and variance calculated from short trajectories as well as in the logistic map (FIG. \ref{fig:tent_curve}).

We calculated $b$ using this equation in the same way as for the logistic map (see Appendix), and the result is shown in the column "Prediction" in TABLE \ref{tab:results}, which well-reproduced the value from the "Numerical” column.

% We calculated $b,\mbox{var}(M),\mbox{var}(V)$ using this equation in the same way as the logistic map[ref:supp], and the result is shown in the column of "Prediction" in the TABLE \ref{tab:results}, and well reproduced the value from "Numerical".
\begin{figure}[H]
 \begin{center}
  \includegraphics[width=86mm]{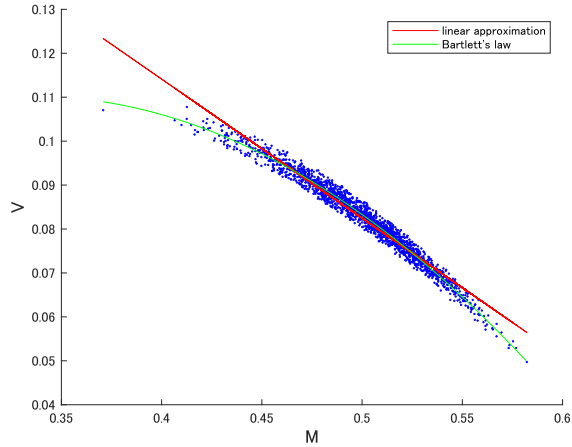}
 \end{center}
 \caption{The typical temporal $M-V$ relationship in the tent map. Here, $\mu = 2$, the number of time steps to calculate the temporal mean and variance is 100, and the number of plots is 2,000. The red and green lines represent the linear and Bartlett's law approximations, respectively.}
 \label{fig:tent_curve}
\end{figure}

The quadratic relationship between the sample mean and variance in general was proposed by Bartlett \cite{Bartlett1936}, and described as follows,
%The quadratic relationship between the temporal mean and variance derived for the logistic map and tent map was proposed by Bartlett \cite{Bartlett1936} and can be generally described as follows:
\begin{eqnarray*}
    V = pM + qM^2.
\end{eqnarray*}
Sometimes, this relationship has been compared to TL \cite{Tayloretal1978, RichardD.Routledge1991}, and, recently, it was called Bartlett's law \cite{Cohen2019}.

% The quadratic relationship between the temporal mean and variance derived for the logistic map and the tent map was proposed by Bartlett \cite{Bartlett1936}, and generally described as follows,
% \begin{eqnarray*}
%     V = pM + qM^2.
% \end{eqnarray*}
% Sometimes, this relationship was compared to TL \cite{Tayloretal1978, RichardD.Routledge1991} and recently it was called Bartlett's law \cite{Cohen2019}.
\subsubsection{Hassell model}

\begin{figure}
 \begin{center}
  \includegraphics[width=86mm]{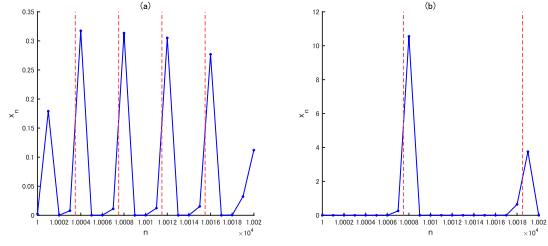}
 \end{center}
 \caption{Chunk structures of the orbits of the Hassell model and Ricker model. (a) Hassell model with $\kappa = 10$, $\nu = 12$, $\lambda = 100$. (b) Ricker model with $r = 5$.}
 \label{fig:chunk_structure}
\end{figure}

The trajectory of the Hassell model in a given parameter region has a chunk structure, in which the value monotonically increases, and, finally, it abruptly decreased to $\sim 0$ to start the next chunk (FIG. \ref{fig:chunk_structure}).
We assume that this chunk structures is responsible for the $M-V$ relationship in the Hassell model. In order to check the importance of the chunk structures, we reshuffled the temporal ensembles, retaining the chunk structure to build a surrogate ensemble (see Appendix). The result from the surrogate is shown in the column "Prediction" in TABLE \ref{tab:results}. In Table I, the prediction values are well-reproduced by the "Numerical" values.

% The trajectory of the Hassell model in a given  parameter region has a chunk structure, in which the value monotonically increased and at last it abruptly decreased to $\sim 0$ to start the next chunk (FIG. \ref{fig:chunk_structure}).
% We assume that this chunk structures is responsible for the $M-V$ relationship in  the Hassell model. In order to check the importance of chunk structures, we reshuffled the temporal ensembles retaining the chunk structure to build a surrogate ensemble [ref:supp]. The result from the surrogate is shown in the column of "Prediction" in the TABLE \ref{tab:results}. In Table I,  Prediction values are well reproduced by the  "Numerical" values.

\subsubsection{Ricker model}

The time trajectory of the Ricker model in this parameter region also has a chunk structure, but, unlike the Hassell model, the temporal ensemble did not show a correlation in $M-V$ at all, and $\mbox{var}(M)$ is much smaller than the spatial ensemble.

First, we constructed the surrogate ensemble in the same way as the Hassell model. The result is shown in the upper row in the column "Prediction" in TABLE \ref{tab:results}. This did not well-reproduce the result from the "Numerical" column, except for $\mbox{var}(V)$.

We found that the discrepancy in $\mbox{var}(M)$ is the direct consequence of the form of the time evolution function, and we analytically showed that the time average depends only on the first term and the $(N+1)$th term;

% The time trajectory of the Ricker model in this parameter region also has a chunk structure, but unlike the Hassell model, the temporal ensemble did not show correlation in $M-V$ at all, and $\mbox{var}(M)$ is much smaller than the spatial ensemble.

% First, we constructed the surrogate ensemble in the same way as the Hassell model. The result is shown in the upper row in the column of "Prediction" in the TABLE. \ref{tab:results}. This did not well reproduce the result from "Numerical", except for $\mbox{var}(V)$.

% We found the discrepancy in $\mbox{var}(M)$ is the direct consequence from the form of the time evolution function, and we analytically showed that the time average depends only on the first term and the $(N+1)$th term, 

\begin{equation}
M = \dfrac{1}{Nr}(\log x_{1} -\log x_{N + 1}) + 1,
\label{eq:RickerM}
\end{equation}
which caused the decorrelation between the mean and variance. This relationship can be generalized to the multi-variable stochastic Ricker model used in \cite{Kilpatrick2003} (see Appedix). 

Based on this, we constructed the surrogate ensemble by random sampling two values from the stationary distribution and applying Eq.\ref{eq:RickerM}. The result is shown in the lower row in the column "Prediction" in TABLE \ref{tab:results}, and it well-reproduced the values in the "Numerical" column.

% Based on this, we constructed surrogate ensemble by random sampling two values from the stationary distribution and applying Eq.\ref{eq:RickerM}. The result is shown in the lower row in the column of "Prediction" in the TABLE \ref{tab:results}, and it well reproduced the value of "Numerical".

\begin{table}[]
\scalebox{0.85}[0.85]{\begin{tabular}{lllllllllllll}
& & \multicolumn{3}{c}{$b = \mbox{cov}(M, V)/\mbox{var}(M)$} & & \multicolumn{3}{c}{$\mbox{var}(M)$} & & \multicolumn{3}{c}{$\mbox{var}(V)$} \\ 
\cline{3-5} \cline{7-9} \cline{11-13} 
Model & Type  & Prediction &  & Numerical &  & Prediction &  & Numerical &  & Prediction &  & Numerical  \\ 
\hline
\begin{tabular}[c]{@{}l@{}}Logistic\\ ($r=3.9$)\end{tabular} 
& Spatial & 
\begin{tabular}[c]{@{}l@{}}-0.056207\\ (-0.056260\\ -0.056149)\end{tabular} &  & \begin{tabular}[c]{@{}l@{}}-0.056188\\ (-0.059249\\ -0.053120)\end{tabular} &  & \begin{tabular}[c]{@{}l@{}}$8.9519*10^{-6}$\\ ($8.9508*10^{-6}$\\ $8.9532*10^{-6}$)\end{tabular} &  & \begin{tabular}[c]{@{}l@{}}$8.9753*10^{-6}$\\ ($8.6127*10^{-6}$\\ $9.2578*10^{-6}$)\end{tabular} &  & \begin{tabular}[c]{@{}l@{}}$4.6849*10^{-7}$\\ ($4.6844*10^{-7}$\\ $4.6854*10^{-7}$)\end{tabular} &  & \begin{tabular}[c]{@{}l@{}}$4.6934*10^{-7}$\\ ($4.5432*10^{-7}$\\ $4.8350*10^{-7}$)\end{tabular} \\
& & &  &  &  &  &  &  &  &  &  & \\
& Temporal & 
\begin{tabular}[c]{@{}l@{}}-0.44142\\ (-0.44147\\ -0.44136)\end{tabular} & & \begin{tabular}[c]{@{}l@{}}-0.44132\\ (-0.44151\\ -0.44113)\end{tabular}    &  & \ &  & \begin{tabular}[c]{@{}l@{}}$1.9021*10^{-6}$\\ ($1.8511*10^{-6}$\\ $1.9562*10^{-6}$)\end{tabular} &  &  &  & \begin{tabular}[c]{@{}l@{}}$3.7065*10^{-7}$\\ ($3.6032*10^{-7}$\\ $3.8131*10^{-7}$)\end{tabular} \\ 
\hline
\begin{tabular}[c]{@{}l@{}}Tent\\ ($\mu=1.9$)\end{tabular}  
& Spatial & 
\begin{tabular}[c]{@{}l@{}}-0.027104\\ (-0.027151\\ -0.027060)\end{tabular}  &  & \begin{tabular}[c]{@{}l@{}}-0.026905\\ (-0.030327\\ -0.023764)\end{tabular} &  & \begin{tabular}[c]{@{}l@{}}$5.5870*10^{-6}$\\ ($5.5859*10^{-6}$\\ $5.5880*10^{-6}$)\end{tabular} &  & \begin{tabular}[c]{@{}l@{}}$5.5960*10^{-6}$\\ ($5.4148*10^{-6}$\\ $5.7732*10^{-6}$)\end{tabular} &  & \begin{tabular}[c]{@{}l@{}}$2.7472*10^{-7}$\\ ($2.7466*10^{-7}$\\ $2.7479*10^{-7}$)\end{tabular} &  & \begin{tabular}[c]{@{}l@{}}$2.7416*10^{-7}$\\ ($2.6625*10^{-7}$\\ $2.8283*10^{-7}$)\end{tabular} \\
& &  &  & &  & &  &  &  &   &  &  \\
& Temporal & 
\begin{tabular}[c]{@{}l@{}}-0.45362\\ (-0.45367\\ -0.45357)\end{tabular}  &  & \begin{tabular}[c]{@{}l@{}}-0.45348\\ (-0.45367\\ -0.45331)\end{tabular} &  &  &  & \begin{tabular}[c]{@{}l@{}}$1.5664*10^{-6}$\\ ($1.5258*10^{-6}$\\ $1.6127*10^{-6}$)\end{tabular} &  &  &  & \begin{tabular}[c]{@{}l@{}}$3.2210*10^{-7}$\\ ($3.1396*10^{-7}$\\ $3.3160*10^{-7}$)\end{tabular} \\ 
\hline
\begin{tabular}[c]{@{}l@{}}Hassell\\ ($\kappa = 10$,\\ $\nu = 12$,\\ $\lambda=100$)\end{tabular}
& Spatial & 
\begin{tabular}[c]{@{}l@{}}0.14187\\ (0.14186\\ 0.14189)\end{tabular} &  & \begin{tabular}[c]{@{}l@{}}0.14186\\ (0.14068\\ 0.14285)\end{tabular}  &  & \begin{tabular}[c]{@{}l@{}}$1.1158*10^{-6}$\\ ($1.1156*10^{-6}$\\ $1.1161*10^{-6}$)\end{tabular} &  & \begin{tabular}[c]{@{}l@{}}$1.1177*10^{-6}$\\ ($1.0737*10^{-6}$\\ $1.1576*10^{-6}$)\end{tabular} &  & \begin{tabular}[c]{@{}l@{}}$2.7672*10^{-8}$\\ ($2.7665*10^{-8}$\\ $2.7680*10^{-8}$)\end{tabular} &  & \begin{tabular}[c]{@{}l@{}}$2.7719*10^{-8}$\\ ($2.6626*10^{-8}$\\ $2.8626*10^{-8}$)\end{tabular} \\ 
&  &  &  &  &  &   &  &   &  &   &  &  \\
& Temporal &
\begin{tabular}[c]{@{}l@{}}0.43951\\ (0.43690\\ 0.44229)\end{tabular}  &  & \begin{tabular}[c]{@{}l@{}}0.58675\\ (0.58500\\ 0.58821)\end{tabular}  &  & \begin{tabular}[c]{@{}l@{}}$5.0776*10^{-8}$\\ ($4.9191*10^{-8}$\\ $5.2334*10^{-8}$)\end{tabular}  &  & \begin{tabular}[c]{@{}l@{}}$4.1391*10^{-8}$\\ $(4.0103*10^{-8}$\\ $4.2418*10^{-8}$)\end{tabular} &  & \begin{tabular}[c]{@{}l@{}}$1.1044*10^{-8}$\\ ($1.0734*10^{-8}$\\ $1.1336*10^{-8}$)\end{tabular} &  & \begin{tabular}[c]{@{}l@{}}$1.4551*10^{-8}$\\ ($1.4133*10^{-8}$\\ $1.4905*10^{-8}$)\end{tabular} \\ 
\hline
\begin{tabular}[c]{@{}l@{}}Ricker\\ ($r=5$)\end{tabular}  
& Spatial  & 
\begin{tabular}[c]{@{}l@{}}7.2162\\ (7.2151\\ 7.2172)\end{tabular}  &  & \begin{tabular}[c]{@{}l@{}}7.2193\\ (7.1719\\ 7.2508)\end{tabular}  &  & \begin{tabular}[c]{@{}l@{}}$6.7806*10^{-4}$\\ ($6.7791*10^{-4}$\\ $6.7820*10^{-4}$)\end{tabular}  &  & \begin{tabular}[c]{@{}l@{}}$6.7666*10^{-4}$\\ ($6.5510*10^{-4}$\\ $7.0472*10^{-4}$)\end{tabular} &  & \begin{tabular}[c]{@{}l@{}}0.038743\\ (0.038725\\ 0.038759)\end{tabular} &  & \begin{tabular}[c]{@{}l@{}}0.038667\\ (0.037303\\ 0.040184)\end{tabular}\\
& & & & & &  &  &  &  &  &  & \\
& Temporal &
\begin{tabular}[c]{@{}l@{}}Chunk\\ 30.484\\ (29.099\\ 31.681)\\ Formula\\ 7.5120\\ (4.1525\\ 11.040)\end{tabular} &  & \begin{tabular}[c]{@{}l@{}}7.3799\\ (3.0874\\ 11.399)\end{tabular} &  & \begin{tabular}[c]{@{}l@{}}Chunk\\ $1.0273*10^{-6}$\\ ($1.0035*10^{-6}$\\ $1.0589*10^{-6}$)\\ Formula\\ $1.5616*10^{-7}$\\ ($1.5217*10^{-7}$\\ $1.5962*10^{-7}$)\end{tabular} &  & \begin{tabular}[c]{@{}l@{}}$1.5599*10^{-7}$\\ ($1.5224*10^{-7}$\\ $1.5934*10^{-7}$)\end{tabular} &  & \begin{tabular}[c]{@{}l@{}}Chunk\\0.0050676\\ (0.0049290\\ 0.0052013)\end{tabular} &  & \begin{tabular}[c]{@{}l@{}}0.0065328\\ (0.0063732\\ 0.0067021)\end{tabular} \\ 
\hline
& &  &  &  &  &  &  &  &  &  &  &  \\
& &  &  &  &  &  &  &  &  &  &  &
\end{tabular}}
\caption{\label{tab:results} The results from the numerical simulations (“Numerical”) and the predicted values from the theoretical models (“Prediction”). In the "Numerical" column, we calculated the indices from 10,000 distinct trajectories. In the "Prediction" column, we calculated the indices from the theoretical models of each map. These results are described with a median and 95\% CI calculated from 200 repeated calculations (95\% CI is given below the associated median value). }
\end{table}

\section{Discussion and Conclusion}

In summary, we have investigated STL and TTL in several chaotic dynamical systems and analyzed the mechanisms of Taylor's law. STL originated from the skewness of the stationary distribution, which is the same mechanism discussed in \cite{Cohen2015}, and the difference between STL and TTL was derived from the temporal correlation of the state variables. In our case, each trajectory was independent of each other, and the relationship between STL and TTL only depends on the temporal correlation in each trajectory. So, our results were complementary to past research \cite{Saitoh2020,Zhao2019}, which considered the case of dependent trajectories and argued how the dependence between the trajectories, such as the environmental synchrony and density-dependent dispersal, affected the exponent, although the possibility that the temporal correlation might affect the exponent of TTL was pointed out in \cite{Zhao2019}.

We also showed that the TTL was well-approximated based on the characteristic temporal structure of each map. In the Hassell model, the temporal structure over several time steps, chunk structures, mostly contributed to the $M-V$ relationships, while, in the logistic map and the tent map, the relationships between the present value and the value of the next time step, which are expressed in the time-evolution equations, strongly influenced the mean-variance relationships. Interestingly, while the trajectories from the Ricker model have chunk structures similar to those of the Hassell model, the TTL cannot be well approximated by the chunk structure due to the presence of the relationship in the time-evolution equation, as found in the logistic map and the tent map. 

The quadratic relation obtained for the TTL of the logistic map and tent map is an example of the relationship called Bartlett's law \cite{Bartlett1936}. Our explanation, based on the relationship between the values of consecutive time steps, can provide another explanation for the quadratic relations.
On the other hand, the mechanism of TL in high-dimensional systems may be different, and further research is needed.

\section*{Acknowledgments}
The authors would like to deeply thank Joel E. Cohen and Yuzuru Sato for valuable discussions.

\nocite{*}
\bibliography{ref} %hoge.bibから拡張子を外した名前
\bibliographystyle{unsrt} %参考文献出力スタイル

\pagebreak
\appendix

\section{Appendix}

\subsection{Calculation Details of Prediction and Numerical}

We evaluated the mean-variance relationship in two ways, $Prediction$ and $Numerical$. In $Numerical$, we made 10,000 distinct trajectories starting from different initial conditions randomly chosen from a uniform distribution between 0 and 1, and the length of the trajectories was 20,000. The first 10,000 steps were discarded to eliminate the effects of transient dynamics. Then, we got 10,000 $\times$ 10,000 data sets. To test spatial TL, we calculated 10,000 ensemble average and variance. From this data, we calculated $\mbox{cov}(M,V)$, $\mbox{var}(M)$, and $\mbox{var}(V)$. On the other hand, we calculated 10,000 time average and variance to test temporal TL. From this data, we calculated $\mbox{cov}(M,V)$, $\mbox{var}(M)$, and $\mbox{var}(V)$. In $Prediction$, to confirm the $Numerical$ results in the spatial format, we made the stationary distributions from the 10,000 $\times$ 10,000 data sets and calculated $\mbox{cov}(M,V)$, $\mbox{var}(M)$, and $\mbox{var}(V)$ using Eq.\ref{eq:sample mean}-\ref{eq:skewness}. To confirm the $Numerical$ results in the temporal format, we performed different analyses for each map in $Prediction$. For the logistic map and tent map, we used Eq.\ref{eq:logisticVM} and Eq.\ref{eq:tentVM}. First, we calculated 10,000 pairs of $(M, V)$ in the spatial format and calculated the average of these $M$ and $V$ ($\overline{M}$ and $\overline{V}$). Next, we calculated the straight lines whose slopes were $\left.\dfrac{dV}{dM}\right|_{M=\overline{M}}$ (this quantity corresponds to $b$) passing through $(\overline{M}, \overline{V})$. For the Hassell model, the 10,000 $\times$ 10,000 dataset was shuffled while maintaining the chunk structure. From these new trajectories, we calculated $b=\mbox{cov}(M,V)/\mbox{var}(M)$, $\mbox{var}(M)$, and $\mbox{var}(V)$. For the Ricker model, in addition to using the same analysis as for the Hassell model, another analysis was also performed using Eq.\ref{eq:RickerM}. $\mbox{var}(M)$ was predicted by randomly extracting two values from the stationary distribution and applying the formula Eq.\ref{eq:RickerM}.
We repeated the above procedure 200 times to create a 95 \% confidence interval.

\subsection{Relationship between temporal TL and auto-correlation}

When we set 
\begin{eqnarray*}
 y_i = x_i - E[x_i],
\end{eqnarray*}
then the following equation holds.
\begin{eqnarray}
 E\left[\sum_{i=1}^{N}y_i\right] =0.
 \label{eq:y_mean}
\end{eqnarray}

\begin{eqnarray}
\label{eq:V_app}
  V &=& \dfrac{1}{N-1} \sum_{i=1}^{N}\left( x_i - \dfrac{1}{N}\sum_{j=1}^{N} x_j \right)^2 \nonumber \\
  &=& \dfrac{1}{N-1} \sum_{i=1}^{N}\left( x_i^2 - \dfrac{2}{N}x_i\sum_{j=1}^{N} x_j + \left(\dfrac{1}{N}\sum_{j=1}^{N} x_j\right)^2
  \right)\nonumber\\
  &=& \dfrac{1}{N-1} \left( \sum_{i=1}^{N}x_i^2 - \dfrac{2}{N}\sum_{i=1}^{N}x_i\sum_{j=1}^{N} x_j + \sum_{i=1}^{N}\left(\dfrac{1}{N}\sum_{j=1}^{N} x_j\right)^2
  \right)\nonumber\\
  &=& \dfrac{1}{N-1} \left( \sum_{i=1}^{N}x_i^2 - \dfrac{1}{N}\sum_{i,j=1}^{N}x_i x_j
  \right)
\end{eqnarray}

\begin{eqnarray*}
   \mbox{var}(M) &=& E\left[ \left( \dfrac{1}{N}\sum_{i=1}^{N}x_i -E\left[\dfrac{1}{N}\sum_{i=1}^{N}x_i\right] \right)^2 \right]\\
   &=& \dfrac{1}{N^2} E \left[ \left(\sum_{i=1}^{N}\left(x_i - E\left[x_i\right]\right)\right)^2 \right]\\
   &=& \dfrac{1}{N^2} E \left[ \left(\sum_{i=1}^{N} y_i\right)^2\right] \\
  & = &\dfrac{1}{N^2} E \left[ \sum_{i=1}^{N} y_i^2 + \sum_{i \neq j}^{N} y_i y_j \right] \\
  & = &\dfrac{\mu_2}{N} + \dfrac{1}{N^2}\sum_{\tau = 1}^{N-1}2(N-\tau) R_1(\tau)
\end{eqnarray*}

\begin{eqnarray*}
   \mbox{var}(V) &=& E \left[ \left(\dfrac{1}{N-1}\sum_{i=1}^{N}\left(x_i- \dfrac{1}{N}\sum_{j=1}^{N}x_j\right)^2 - E\left[\dfrac{1}{N-1}\sum_{i=1}^{N}\left(x_i - \dfrac{1}{N}\sum_{j=1}^{N}x_j\right)^2\right] \right)^2 \right]\\
   &=& E \left[ \left(\dfrac{1}{N-1}\sum_{i=1}^{N}\left(y_i- \dfrac{1}{N}\sum_{j=1}^{N}y_j\right)^2 - E\left[\dfrac{1}{N-1}\sum_{i=1}^{N}\left(y_i - \dfrac{1}{N}\sum_{j=1}^{N}y_j\right)^2\right] \right)^2 \right]\\
   &=& E\left[ \left( \dfrac{1}{N-1}\left(\sum_{i=1}^{N}y_i^2 - \dfrac{1}{N}\sum_{i,j=1}^{N}y_i y_j \right) - E\left[ \dfrac{1}{N-1}\left(\sum_{i=1}^{N}y_i^2 - \dfrac{1}{N}\sum_{i,j=1}^{N}y_i y_j \right)\right] \right)^2 \right]\ \ \left(\because Eq.\ref{eq:V_app} \right)\\
   &\simeq& E\left[ \left( \dfrac{1}{N}\sum_{i=1}^{N}\left( y_i^2 - E[y_i^2] \right) \right)^2 \right]\\
   &=& \dfrac{1}{N^2}\sum_{i,j=1}^{N}E\left[ \left( y_i^2 - E[y_i^2] \right) \left( y_j^2 - E[y_j^2] \right)\right]\\
   &=& \dfrac{1}{N^2}\left( \sum_{i=1}^{N}E\left[ \left( y_i^2 - E[y_i^2] \right)^2\right] + \sum_{i\neq j}^{N}E\left[\left( y_i^2 - E[y_i^2] \right) \left( y_j^2 - E[y_j^2] \right)\right] \right)\\
   &=& \dfrac{1}{N}\left(\mu_4 - \mu_2^2 \right) + \dfrac{1}{N^2}\sum_{\tau = 1}^{N-1}2(N-\tau) R_2(\tau)
\end{eqnarray*}

\begin{eqnarray*}
   \mbox{cov}(M,V) &=& E\left[ \left( \dfrac{1}{N}\sum_{i=1}^{N}x_i - E\left[\dfrac{1}{N}\sum_{i=1}^{N}x_i\right]\right)\left( \dfrac{1}{N-1}\sum_{i=1}^{N}\left(x_i - \dfrac{1}{N}\sum_{j=1}^{N}x_j\right)^2 - E\left[\dfrac{1}{N-1}\sum_{i=1}^{N}\left(x_i - \dfrac{1}{N}\sum_{j=1}^{N}x_j\right)^2\right] \right) \right]\\
  &=& \dfrac{1}{N(N-1)}E\left[ \left(\sum_{i=1}^{N}y_i\right) \left( \sum_{i=1}^{N}\left( y_i -\dfrac{1}{N}\sum_{j=1}^{N}y_j \right)^2 - E\left[ \sum_{i=1}^{N}\left( y_i-\dfrac{1}{N}\sum_{j=1}^{N}y_j \right)^2 \right]\right) \right] \\
  & = &\dfrac{1}{N(N-1)}E\left[ \sum_{i=1}^{N}y_i \left( \sum_{i=1}^{N}\left( y_i -\dfrac{1}{N}\sum_{j=1}^{N}y_j \right)^2\right) \right] \left( \because Eq.\ref{eq:y_mean} \right) \\
  & = &\dfrac{1}{N(N-1)}E\left[ \sum_{i=1}^{N}y_i \left( \sum_{i=1}^{N} y_i^2 -\dfrac{1}{N}\sum_{i=1}^{N}y_i\sum_{j=1}^{N}y_j \right) \right]\left( \because Eq.\ref{eq:V_app} \right)\\
  & = &\dfrac{1}{N(N-1)}E\left[ \sum_{i,j=1}^{N}y_i y_j \left(y_j - \dfrac{1}{N}\sum_{k=1}^{N} y_k \right) \right] \\
  & \simeq& \dfrac{1}{N^2}E \left[ \sum_{i,j=1}^{N}y_i y_j^2 \right] \\
  & = &\dfrac{1}{N^2}E \left[ \sum_{i=1}^{N}y_i^3 + \sum_{i\neq j}^{N}y_i y_j^2 \right] \\
  & = &\dfrac{\mu_3}{N} + \dfrac{1}{N^2}\sum_{\tau = 1}^{N-1}(N-\tau)(R_{12}(\tau) + R_{12}(-\tau))
\end{eqnarray*}

\subsection{Bartlett's law of quadratic dynamical systems}

Dynamical systems described in quadratic form:
\begin{eqnarray}
    x_{n+1} = c x_{n}^2 + d x_{n},
    \label{eq:quadraticDS}
\end{eqnarray}
where $c$ and $d$ are constants, satisfy Bartlett's law. We define the sample mean ($M$) and the sample variance ($V$) as follows:

\begin{eqnarray*}
    M &=& \dfrac{1}{N}\sum_{n=1}^N x_{n},\\
    V &=& \dfrac{1}{N-1}\sum_{n=1}^N (x_n - M)^2.
\end{eqnarray*}
Bartlett's law is described as follows:
\begin{eqnarray*}
    V &=& p M + q M^2 .
\end{eqnarray*}
Proof.\\
\begin{eqnarray*}
    x_{n+1} &=& c x_{n}^2 + d x_{n}, \\
            &=& c\left(x_{n} + \dfrac{d}{2c}\right)^2 -\dfrac{d^2}{4c},\\
            &=& c\left\{ (x_{n} - M) + \left(M + \dfrac{d}{2c}\right)\right\}^2 -\dfrac{d^2}{4c},\\
            &=& c(x_{n}-M)^2 + c\left(M + \dfrac{d}{2c}\right)^2 + 2c\left( M + \dfrac{d}{2c}\right)(x_{n}-M) - \dfrac{d^2}{4c}.
\end{eqnarray*}
Taking the sum and dividing by $N$ on both sides of the equation, and if $N$ is large enough:
\begin{eqnarray*}
\begin{split}
    \mbox{(Left side)} &= \dfrac{1}{N}\sum_{n=1}^N x_{n+1}, \\
                     &\simeq \dfrac{1}{N}\sum_{n=1}^N x_{n}, \\
                     &= M.
\end{split}
\end{eqnarray*}
On the other hand:
\begin{eqnarray*}
    \mbox{(Right side)} &=& \dfrac{1}{N} \sum_{n=1}^N \left\{c(x_{n}-M)^2 + c\left(M + \dfrac{d}{2c}\right)^2 \right\} + 2c\left(M + \dfrac{d}{2c}\right)\dfrac{1}{N}\sum_{n=1}^N (x_{n}-M) - \dfrac{d^2}{4c} , \\
    &\simeq& c V + c\left(M + \dfrac{d}{2c}\right)^2 - \dfrac{d^2}{4c}.
\end{eqnarray*}
Therefore, we obtain:
\begin{eqnarray*}
    M &=&  c V + c\left(M + \dfrac{d}{2c}\right)^2 - \dfrac{d^2}{4c},\\
    \Leftrightarrow V &=& \dfrac{1-d}{c} M - M^2.
\end{eqnarray*}

\subsection{Bartlett's law of tent map}
Tent map (Eq.\ref{eq:tent}) satisfies Bartlett's law with $p = \dfrac{\mu}{\mu + 1}, q = -1$.
\begin{eqnarray}
x_{n+1} &=& \dfrac{\mu}{2} - \mu \left| x_n - \dfrac{1}{2} \right|.
\label{eq:tent}
\end{eqnarray}
Proof.
\begin{eqnarray*}
    x_{n+1} &=& \dfrac{\mu}{2} - \mu \left| x_n - \dfrac{1}{2} \right|, \\
    \Leftrightarrow \left| x_n - \dfrac{1}{2} \right| &=& \dfrac{1}{2} - \dfrac{1}{\mu} x_{n+1}.
\end{eqnarray*}

By squaring both sides, we obtain,
\begin{eqnarray*}
    \left( x_n - \dfrac{1}{2} \right)^2 &=& \left( \dfrac{1}{2} - \dfrac{1}{\mu} x_{n+1} \right)^2.
\end{eqnarray*}
\begin{eqnarray*}
 \mbox{(Left side)} &=& \left\{ (x_n - M) + \left(M - \dfrac{1}{2} \right) \right\}^2,\\
 &=& \left(x_{n} - M \right)^2 + \left( M - \dfrac{1}{2} \right)^2 + 2(x_{n} - M)\left( M - \dfrac{1}{2} \right).
\end{eqnarray*}
\begin{eqnarray*}
 \mbox{(Right side)} &=& \dfrac{1}{\mu ^2}\left\{ (x_{n+1} - \mbox{M}) + \left( \mbox{M} - \dfrac{\mu}{2} \right) \right\}^2,\\
 &=& \dfrac{1}{\mu ^2}\left\{ (x_{n+1} - \mbox{M})^2 + \left( \mbox{M} - \dfrac{\mu}{2} \right)^2 + 2\left( \mbox{M} - \dfrac{\mu}{2} \right)(x_{n+1} - \mbox{M}) \right\}.
\end{eqnarray*}

Taking the sum of both sides of the equation and dividing both sides by $N$, we obtain the following equations:

%  Taking the sum of both sides of the equation and dividing both sides by $N$, we obtain the following equations:
\begin{eqnarray*}
    \mbox{(Left side)} &=&\dfrac{1}{N}\sum_{n=1}^N \left\{ \left(x_{n} - \mbox{M}\right)^2 + \left( \mbox{M} - \dfrac{1}{2} \right)^2+ 2(x_{n} - \mbox{M})\left( \mbox{M} - \dfrac{1}{2} \right) \right\}, \\
    % &= \dfrac{1}{N}\sum_{n=1}^N \left(x_{n} - \mbox{M}\right)^2 + \dfrac{1}{N}\sum_{n=1}^N \left(\mbox{M} - \dfrac{1}{2}\right)^2 \\
    % & \quad \quad \quad \quad \quad \quad + 2\left( \mbox{M} - \dfrac{1}{2}\right)\dfrac{1}{N} \sum_{n=1}^N ( x_{n} - \mbox{M}), \\
    &\simeq& \mbox{V} + \left(\mbox{M} - \dfrac{1}{2}\right)^2.
\end{eqnarray*}

\begin{eqnarray*}
    \mbox{(Right side)} &=& \dfrac{1}{\mu ^2} \dfrac{1}{N}\sum_{n=1}^N \left( x_{n+1} - M\right)^2 + \dfrac{1}{\mu ^2} \dfrac{1}{N}\sum_{n=1}^N \left(M - \dfrac{\mu}{2}\right)^2 + \dfrac{2}{\mu ^2}\left( M - \dfrac{\mu}{2} \right)\dfrac{1}{N}\sum_{n=1}^N (x_{n+1} - M).
\end{eqnarray*}
If $N$ is large enough,
\begin{eqnarray*}
     \dfrac{1}{N}\sum_{n=1}^N (x_{n+1} - M)^2 &\simeq& \dfrac{1}{N-1}\sum_{n=1}^N (x_{n} - M)^2, \\
     &=& V,
\end{eqnarray*}
\begin{eqnarray*}
     \dfrac{1}{N}\sum_{n=1}^N (x_{n+1} - M) &\simeq& \dfrac{1}{N}\sum_{n=1}^N (x_{n} - M), \\
     &=& 0,
\end{eqnarray*}
we obtain:
\begin{eqnarray}
    V + \left(M - \dfrac{1}{2}\right)^2 &=& \dfrac{1}{\mu ^2}V + \dfrac{1}{\mu ^2}\left( M - \dfrac{\mu}{2} \right)^2, \nonumber \\
    \Leftrightarrow V &=& \dfrac{\mu}{\mu + 1}M - M^2.
    \label{eq:BLoftent}
\end{eqnarray}

\subsection{Temporal Mean of Ricker model}

We applied $\log$ to each side of $x_{n+1} = x_{n}\exp \{r(1-x_{n})\}$ and obtained the following equation.

\begin{equation}
 \log x_{n+1} = \log x_{n} + r(1-x_n)
\end{equation}

Using this equation iteratively, we obtain:
\begin{eqnarray*}
\begin{split}
    \log(x_{N+1}) &= \log(x_{N}) + r(1-x_N)\\
    & = \log(x_{N-1}) + r(1-x_{N-1}) + r(1-x_N) \\
    & ... \\
    &= \log(x_{1}) + \sum_{n=1}^N r(1-x_{n}),\\
    \Leftrightarrow M &=\dfrac1{Nr}(\log(x_{1})-\log(x_{N+1}))+1.\\
\end{split}
\end{eqnarray*}

In a similar way, we can obtain the temporal mean of the multi-variable stochastic Ricker model:
\begin{equation}
    x^{i}_{n+1} = x^{i}_{n}\exp \left\{r^{i}\left(1-\dfrac{x^{i}_{n}+\sum_{i\neq j} \alpha^{ij}x^{j}_{n}}{K^{i}}\right)+\epsilon^{i}_{n}\right\}.
    \label{eq:Ricker_general_2}
\end{equation}
Considering the log of the each side of Eq.\ref{eq:Ricker_general_2}, we get:
\begin{equation}
    \log (x^{i}_{n+1}) = \log (x^{i}_{n}) + r^{i}\left(1-\dfrac{x^{i}_{n}+\sum_{i\neq j} \alpha^{ij}x^{j}_{n}}{K^{i}}\right)+\epsilon^{i}_{n}.
    \label{eq:Ricker_general_log}
\end{equation}
Using this equation iteratively, we obtain:
\begin{eqnarray*}
    \log (\bm{x}_{n+1}) = \log (\bm{x}_{1}) + N\bm{r} - \dfrac{r^i}{K^i}A\sum^{N}_{n=1}\bm{x}_n + \sum^{N}_{n=1}\bm{\epsilon}_{n}, \\
    \Leftrightarrow \dfrac{r^i}{K^i}A\sum^{N}_{n=1}\bm{x}_n = \log (\bm{x}_{1}) -  \log (\bm{x}_{n+1}) + N\bm{r} + \sum^{N}_{n=1}\bm{\epsilon}_{n},
\end{eqnarray*}
,where $(\bm{x}_{n})^{i} = x^{i}_{n}, (\bm{\epsilon}_{n})^{i} = \epsilon^{i}_{n}, (A)^{ij} =\alpha^{ij} (i\neq j),(A)^{ii} = 1$,and $\bm{r}^i = r^i$. Therefore, we can get the following formula:
\begin{eqnarray*}
    M^i &=& \dfrac{1}{N}\sum^{N}_{n=1} \bm{x}_n^i \\
        &=& \dfrac{K^{i}}{Nr^{i}}A^{-1}\left\{\log \bm{x}_1 -\log \bm{x}_{N + 1} + N\bm{r}+\sum_{n=1}^{N}\bm{\epsilon}_{n}\right\}^{i}
\end{eqnarray*}

\end{document}